# Solitons supported by singular spatial modulation of the Kerr nonlinearity

Olga V. Borovkova,[1] Valery E. Lobanov,[1] and Boris A. Malomed[2,1]

[1]*ICFO-Institut de Ciencies Fotoniques, and Universitat Politecnica de Catalunya, Mediterranean Technology Park, 08860 Castelldefels (Barcelona), Spain*

[2]*Department of Physical Electronics, School of Electrical Engineering, Faculty of Engineering, Tel Aviv University, Tel Aviv 69978, Israel*

**Abstract**

We introduce a setting based on the one-dimensional (1D) nonlinear Schrödinger equation (NLSE) with the self-focusing (SF) cubic term modulated by a singular function of the coordinate, $|x|^{-\alpha}$. It may be additionally combined with the uniform self-defocusing (SDF) nonlinear background, and with a similar singular repulsive linear potential. The setting, which can be implemented in optics and BEC, aims to extend the general analysis of the existence and stability of solitons in NLSEs. Results for fundamental solitons are obtained analytically and verified numerically. The solitons feature a *quasi-cuspon* shape, with the second derivative diverging at the center, and are stable in the entire existence range, which is $0 \leq \alpha < 1$. Dipole (odd) solitons are found too. They are unstable in the infinite domain, but stable in the semi-infinite one. In the presence of the SDF background, there are two subfamilies of fundamental solitons, one stable and one unstable, which exist together above a threshold value of the norm (total power of the soliton). The system which additionally includes the singular repulsive linear potential emulates solitons in a uniform space of the *fractional dimension*, $0 < D \leq 1$. A 2D extension of the system, based on the quadratic ($\chi^{(2)}$) nonlinearity, is formulated too.

## I. Introduction and the model

Solitons and their stability in nonlinear Schrödinger and Gross-Pitaevskii equations (NLSEs/GPEs) in spaces of different dimensions $D$ and with various types of the nonlinearity is a general problem of mathematical physics, with vast applications to nonlinear optics, plasmas, Bose-Einstein condensates (BECs), etc. [1-4]. A fundamental restriction on the stability of solitons supported by the self-focusing (SF) nonlinear term $|q|^{2s}q$, where $q$ is the corresponding wave field, is imposed by the onset of the *critical collapse*, according to the Talanov's criterion [1]: $sD<2$.

Recently, the studies of solitons in NLSEs have been expanded by considering various settings with spatially modulated nonlinearity, see original works [5-15] and review [16]. Available experimental techniques allow one to engineer modulation profiles in optics by doping waveguides with resonant atoms [17], or filling voids in photonic crystals by a material with the linear refractive index matched to that of the host medium but a different value of the Kerr coefficient [9,11,16]. In BEC, one may use, for the same purpose, the Feshbach resonance controlled by spatially inhomogeneous fields [5-7,10,12-14]. In particular, the magnetic field may be properly shaped by sets of ferromagnetic films [18].

While recent works on this topic were chiefly focused on regular nonlinearity-modulation patterns, traditional limits of the studies of localized modes may be transcended by the consideration of *singular modulations*. Recently, it has been demonstrated that the use of the *self-defocusing* (SDF) cubic nonlinearity with the local strength diverging at $|x|\to\infty$, or at finite values of coordinate $x$, allows one to support *bright solitons* in the absence of the SF, without the help of any linear potential [19,20]. Similar results were obtained in a system which can be transformed into the same form, with a constant coefficient of the cubic SDF nonlinearity and a spatially modulated diffraction coefficient vanishing at $|x|\to\infty$ [21]. In optics and BEC alike, singular nonlinearity patterns can be created by using modulation profiles with an exact resonance attained at designated singular point(s) [19,20,22].

The objective of the present work is to study fundamental and higher-order one-dimensional (1D) solitons supported by the cubic SF nonlinearity subject to the basic singular modulation. In the most general form, the model is represented by the following NLSE/GPE:

$$iq_z = -\frac{1}{2}q_{xx} + \left(\sigma - |x|^{-\alpha}\right)|q|^2 q + \beta |x|^{-\alpha} q. \qquad (1)$$

Here $z$ is the propagation distance in the NLSE, or time in the GPE, $\alpha > 0$ is the singularity power, the coefficient in front of the SF term is scaled to be 1, and $\sigma \geq 0$ takes into account a possibility that the singular SF nonlinearity may be embedded into the SDF back-ground. As concerns the linear singular potential with coefficient $\beta$, it is shown below that it may be used to emulate the NLSE in the space of a *fractional dimension*, $0 < D \leq 1$, although we chiefly focus on the case of $\beta = 0$.

Our analysis shows that stable (non-collapsing) solitons exist at $0 \leq \alpha < 1$, i.e., for a relatively weak singularity. The numerical results demonstrate that the solitons are also stable against replacement of the singular modulation profile by a regularized one, i.e., the true singularity is not necessary for the creation of the predicted solitons in the experiment. Further, stable solitons found below are not specifically narrow, hence Eq. (1), derived in the standard paraxial approximation, is valid for modeling the light transmission in optical waveguides. Because we aim to describe the stable solitons, rather than the complex dynamics of collapsing fields, Eq. (1) is reliable too in the context of BEC (the mean-field equation might not be relevant as a dynamical model of the collapse).

The analysis of solitons in the model based on Eq. (1) is presented below as follows. Analytical results (both exact and approximate ones) for fundamental solitons are collected in Section II. Numerical findings, which corroborate the analytical predictions for the fundamental solitons, and exhibit new results for dipoles, are reported in Section III. The paper is concluded by Section IV, where we also put forward an extension of the singular-modulation problem for 2D, using the quadratic (second-harmonic-generating) nonlinearity, instead of the cubic one.

## II. Analytical results for fundamental solitons.

We look for stationary solutions to Eq. (1) as $q = \exp(ibz) w(x)$, with propagation constant $b > 0$ (or chemical potential $-b$ in the case of the GPE), and real function $w(x)$ obeying the following equation:

$$bw = \frac{1}{2} w'' + \left( |x|^{-\alpha} - \sigma \right) w^3 - \beta |x|^{-\alpha} w . \qquad (2)$$

Using an obvious rescaling, one can immediately derive an exact relation for the total norm of solitons in the basic case of $\sigma = \beta = 0$:

$$U(b) \equiv \int_{-\infty}^{+\infty} |q(x)|^2 \, dx = b^{(1-\alpha)/2} U(b=1) , \qquad (3)$$

hence the solitons satisfy the Vakhitov-Kolokolov (VK) stability criterion [2,3], $dU/dk > 0$, at $\alpha < 1$, and are definitely unstable at $\alpha \geq 1$ (if they exist in the latter case, see below). Note that this conclusion does not depend on the type of the solitons, pertaining to fundamental and higher-order ones alike.

It is interesting to note that Eq. (2) with $\sigma = b = 0$ may emulate the radial version of the cubic NLSE in the uniform space of the *fractional dimension*,

$$D = 2(1-\alpha)/(2-\alpha), \qquad (4)$$

with the corresponding propagation constant $\beta$, if coordinate $x$ is mapped into $r \equiv 2(2-\alpha)^{-1}|x|^{(2-\alpha)/2}$. Indeed, the accordingly transformed Eq. (2) takes the form of

$$\beta w = \frac{1}{2}\left(\frac{d^2 w}{dr^2} + \frac{D-1}{r}\frac{dw}{dr}\right) + w^3. \qquad (5)$$

For the singularity power $\alpha$ growing from 0 to 1, effective dimension (4) decreases from 1 to 0. Solitons in the space of the fractional dimension correspond to $\beta > 0$ in Eq. (5). In terms of the underlying equation (1), $\beta > 0$ represents a *repulsive* singular linear potential, which then competes with the self-attractive singular nonlinearity. The possibility to use the present system for emulating the fractional-dimension space is relevant in the connection to the currently active topic of "simulating" complex physical media by means of relatively simple matter-wave settings (or photonic ones, as the above discussion suggests) [23].

The shape of fundamental solitons near the singular point can be found by the straightforward expansion of Eq. (2) at $x \to 0$:

$$w(x) = w_0\left[1 - 2\frac{w_0^2 - \beta}{(1-\alpha)(2-\alpha)}|x|^{2-\alpha} + (b+\sigma w_0^2)x^2 + O(|x|^{2(2-\alpha)})\right], \qquad (6)$$

where $w_0$ is the soliton's amplitude attained at $x=0$, and $\beta < w_0^2$ is assumed (otherwise, the soliton does not exist). This soliton may be classified as a "quasi-cuspon": while ordinary cuspons are characterized by a finite amplitude and a diverging first derivative at $x \to 0$

(see, e.g., Ref. [24]), Eq. (6) features $w'(0)=0$ and $w''(x)$ diverging at $x \to 0$. The fundamental soliton does not exist at $\alpha > 1$, as in that case expansion (5) represents a non-soliton solution, with a local *minimum*, rather than maximum, at $x=0$. Precisely at $\alpha=1$, Eq. (6) is replaced by the following expansion:

$$w(x) \approx w_0 \left[ 1 + 2(w_0^2 - \beta)|x|\ln(1/|x|) \right],$$

also with a local minimum at $x=0$ (for $\beta < w_0^2$), hence the fundamental soliton does not exist in this case either.

The soliton family as a whole may be approximated by the Gaussian variational *ansatz*,

$$w^2 = \frac{U}{\sqrt{\pi}W} \exp\left( -\frac{x^2}{W^2} \right), \tag{7}$$

with width $W$ and norm $U$ (the total power, in terms of optics) defined as per Eq. (3). The substitution of the ansatz into the Lagrangian of Eq. (2) with $\beta=0$, $L = \int_{-\infty}^{+\infty} \left[ bw^2 + (1/2)(w')^2 + (1/2)(\sigma - |x|^{-\alpha})w^4 \right] dx$, and the subsequent integration yields

$$L = \left( b + \frac{1}{4W^2} \right) U + \frac{1}{2\sqrt{2\pi}W} \left[ \sqrt{\pi}\sigma - \Gamma\left( \frac{1-\alpha}{2} \right) \left( \frac{\sqrt{2}}{W} \right)^\alpha \right] U^2,$$

where $\Gamma$ is the Gamma-function. The variational equations following from here, $\partial L/\partial U = \partial L/\partial W = 0$, predict the existence of two subfamilies of the solitons at $\sigma > 0$, one VK-stable (with $dU/dk > 0$), and another unstable, above a threshold value of the norm, which can be found analytically:

$$U_{\text{thr}} = \pi^{(1/\alpha+1)/2} \left( \frac{1}{\alpha} - 1 \right) \sigma^{1/\alpha-1} \left[ (1-\alpha^2) \Gamma\left( \frac{1-\alpha}{2} \right) \right]^{-1/\alpha}. \tag{8}$$

Because we consider the case of $\alpha < 1$, it follows from Eq. (8) that the threshold vanishes along with the SDF background, $\sigma = 0$. In the same case, the variational approxima-tion yields explicit results for the norm, width, and amplitude of the fundamental soliton, as functions of propagation constant $b$:

$$U = \frac{\pi [8/(3-\alpha)]^{(1-\alpha)/2} b^{(1-\alpha)/2}}{(1+\alpha)^{(1+\alpha)/2} \Gamma((1-\alpha)/2)}, \quad W^2 = \frac{3-\alpha}{4(1+\alpha)b}, \quad A_{\max}^2 = \frac{U}{\sqrt{\pi}W}. \tag{9}$$

As seen from here, the soliton's amplitude falls to zero at $\alpha \to 1$, $A_{\max}^2 \approx (1/2)\sqrt{\pi b}(1-\alpha)$, while its width remains finite, $W^2 \approx 1/(4b)$. This implies that the paraxial approximation, employed in the derivation of Eq. (1) as the model of optical media, remains valid in this limit too.

In the case of the weak singularity, $\alpha \ll 1$, another approximation may be used, based on expansion $|x|^{-\alpha} \approx 1 + \alpha \ln(1/|x|)$. After simple manipulations, the corresponding soliton solution for $\beta = 0$ and $\sigma < 1$ may be reduced to the ordinary NLSE soliton with amplitude $\sqrt{2b}$, effective mass $m \equiv 1$, and central coordinate $\xi$, moving in an effective potential defined by an integral expression, which can be expanded at small $\xi$:

$$\Pi(\xi) = -\frac{\alpha b}{1-\sigma} \int_{-\infty}^{+\infty} \ln\left(\frac{1}{|y|}\right) \operatorname{sech}^4\left(y - \sqrt{2b}\xi\right) dy \approx \frac{\alpha b}{1-\sigma}(-0.83 + 2.46 b \xi^2).$$

Accordingly, the soliton is predicted to perform small oscillations around $\xi = 0$ with squared frequency $\omega^2 \approx 4.92(1-\sigma)^{-1} \alpha b^2$. For instance, at $\sigma = 0$ and $b = 2$ this formula predicts the period of oscillations $T = 2\pi/\omega \approx 9.0$ and $T \approx 3.2$, for $\alpha = 0.025$ and $\alpha = 0.2$, respectively, while direct simulations of Eq. (1) yield, in the same cases, $T \approx 10$ and $T \approx 3$.

To complete the presentation of the analytical results, it is relevant to mention that, in terms of the scaling with respect to $x$, Eq. (1) with $\alpha = 1$ and $\beta = 0$, which, as shown above, does not produce fundamental solitons, is similar to the equation with the delta-functional SF term, that was introduced in Ref. [25]:

$$iq_z = -\frac{1}{2}q_{xx} + [\sigma - \delta(x)]|q|^2 q. \tag{10}$$

It is easy to find that Eq. (10) gives rise to *exact* soliton solutions, $q = \exp(ibz)\sqrt{2b/\sigma}\left[\sinh\left(2\sqrt{2b}(|x|+\xi)\right)\right]^{-1}$, with $\xi$ defined by relation $\sigma \sinh(2\sqrt{2b}\xi) = 2\sqrt{2b}$, and norm $U(b) = \sqrt{1+8b/\sigma} + 1 - 2\sqrt{2b}/\sigma$. However, this $U(b)$ dependence does not satisfy the VK criterion, hence the solitons are unstable. In the absence of the SDF background, $\sigma = 0$, the norm is degenerate, $U \equiv 1$, which suggests an instability too, as in the case of the *Townes solitons*, associated with the case of the *critical collapse* [3]. Indeed, Eq. (10) with $\sigma = 0$ (as well as the quintic NLSE) realizes the critical collapse in one dimension, as shown by exact solutions for collapsing and decaying solitons [26].

### III. Numerical results

The analytical predictions were verified by numerical solutions of Eqs. (1) and (2). For this purpose, the singular modulation function was replaced by a regularized expression, $|x|^{-\alpha} \to (x^2 + \varepsilon^2)^{-\alpha/2}$, with sufficiently small $\varepsilon$. It was found that the fundamental solitons exist and are *completely stable*, in terms of eigenvalues of perturbation modes and direct simulations alike, in the whole interval of $0 \leq \alpha < 1$, in full agreement with the above prediction based one the VK criterion. Examples of the solitons, displayed in Fig. 1, demonstrate, as expected, the drop of the height and steepening of the profile with the increase of $\alpha$ towards 1.

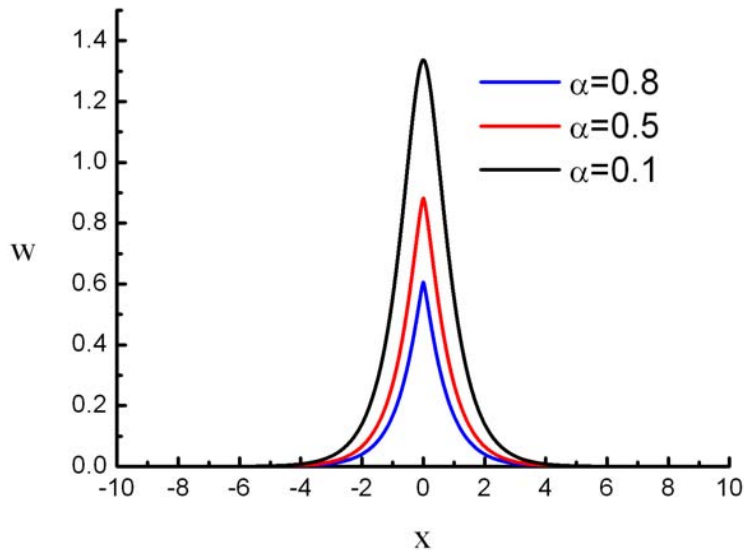

Fig. 1 (Color online) Stable fundamental solitons, found as numerical solutions of Eq. (2) with $\sigma=\beta=0, b=1$ and several values of the singularity power. In this and other figures, all quantities are plotted in the same scaled units in which Eqs. (1) and (2) are written.

In the absence of the SDF background and linear potential ($\sigma=\beta=0$), the soliton family is characterized by dependences of the norm and amplitude on the singularity power, $\alpha$, at $b=1$ [as said above, $b=1$ can be fixed by scaling]. The dependences are displayed in Fig. 2, along with the predictions of the variational approximation given by Eq. (9). Very close agreement between the variational and numerical results is obvious.

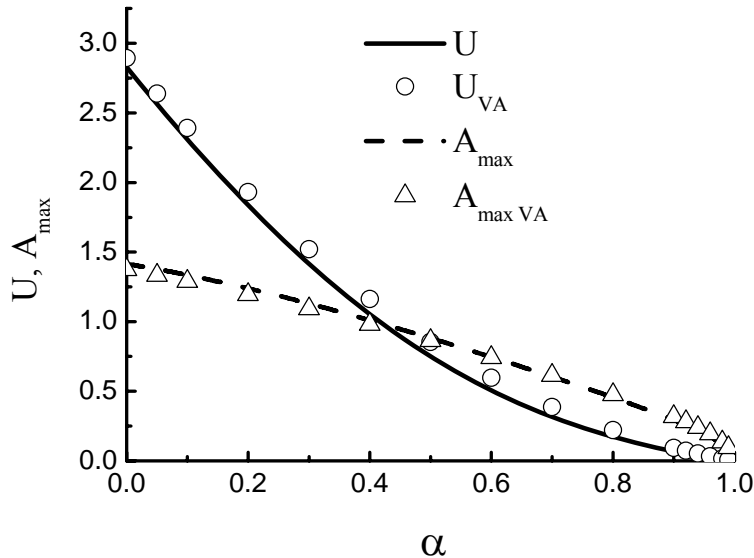

Fig. 2. The norm ($U$) and amplitude of the stable fundamental solitons vs. the singularity power, $\alpha$, for $\sigma=\beta=0$ and the propagation constant scaled to $b=1$, as found from the numerical solution of Eq. (2) (curves), and as predicted by the variational approximation in the form of Eq. (9) ("VA", chains of symbols).

As seen in Fig. 3, in the presence of the SDF background ($\sigma=1, \beta=0$), the numerical solution gives rise to the fundamental solitons at $U > U_{\text{thr}}$, again in agreement with the analytical prediction. In particular, Eq. (8) yields $U_{\text{thr}} \approx 2$ at $\alpha=0.25$. A discrepancy with $U_{\text{thr}} \approx 1.4$ observed in Fig. 3 is explained by the difference of the Gaussian ansatz (7) from the exact (quasi-cuspon) shape of the solitons.

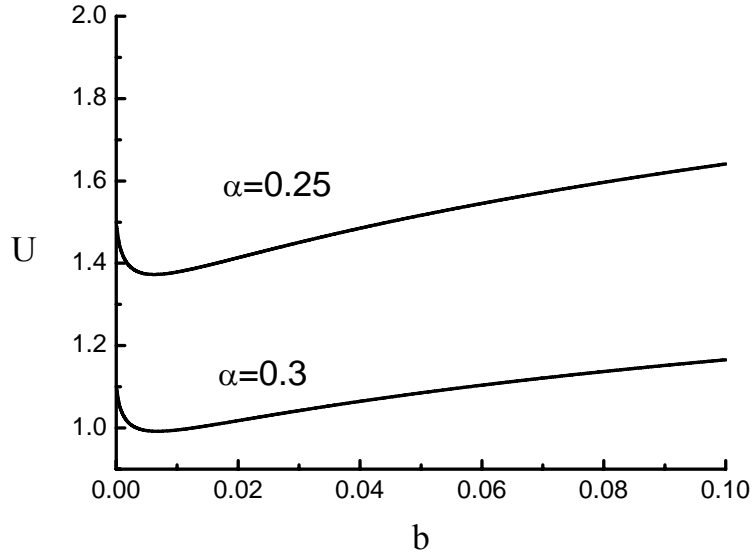

Fig. 3. The numerically found norm of the fundamental solitons vs. the propagation constant, for $\sigma=1$ and $\beta=0$ in Eq. (2). In agreement with the VK criterion, the right and left portions of the family are found to be stable and unstable, respectively.

The system gives rise to higher-order solitons too, including dipoles (odd modes). In particular, the expansion of the dipole solution near $x=0$ yields $w(x)=w_1 x\left[1-2(5-\alpha)^{-1}(4-\alpha)^{-1} w_1^2 |x|^{4-\alpha} +(b/3)x^2\right]+...$ for $\sigma=\beta=0$, where $w_1$ is an arbitrary parameter of the dipole family [cf. Eq. (6)]. As follows from here and is confirmed by the numerical solution of Eq. (2), the dipoles, unlike the fundamental solitons, exist at $\alpha \geq 1$ too, but the above analysis demonstrates that they are VK-unstable in that case. At $\alpha<1$, the dipoles are found to be unstable against spontaneous merger into fundamental solitons, see Fig. 4 (this symmetry-breaking instability mode is ignored by the VK criterion). Indeed, a detailed analysis demonstrates that the single eigenmode of small perturbations accounting for the instability is an even one, with a maximum at $x=0$ (not shown here), in contrast with the odd structure of the dipole.

The dipoles are stabilized in the semi-infinite space, with boundary condition (b.c.) $q(x=0)=0$, which may be realized in BEC, as proposed in Ref. [22], by placing the singular point at a solid surface. This b.c. rules out fundamental solitons, but makes the dipole stable, eliminating the above-mentioned instability eigenmode. In the guided-wave-propagation setting, this b.c. may be realized too – e.g., as a metallic wall of a microwave duct, onto which the nonlinearity-enhancing dopant is deposited. Higher-order modes were found too, but they are completely unstable against the fusion into fundamental solitons.

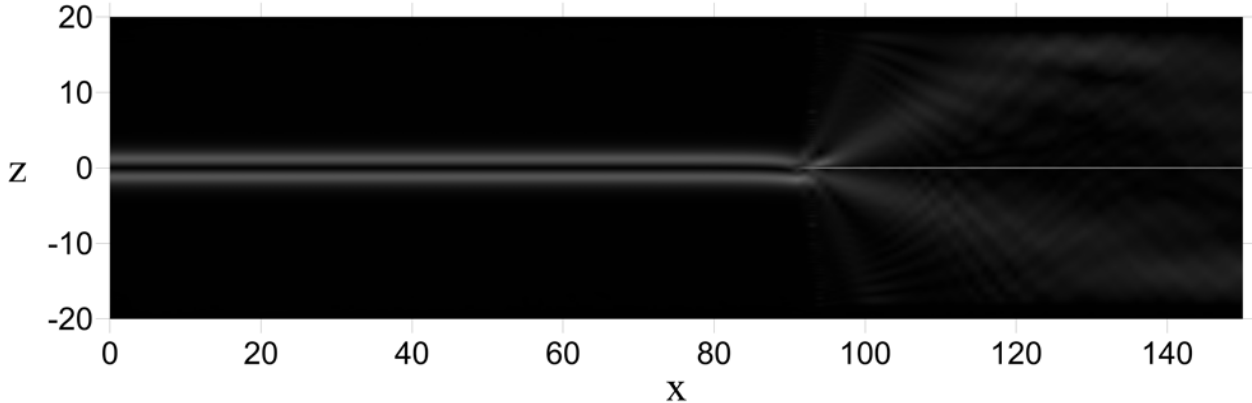

Fig. 4. A contour plot of $|q(x,z)|$ demonstrating the spontaneous transformation of an unstable dipole into a fundamental soliton at $\sigma=\beta=0, \alpha=0.5, b=1$. The respective instability growth rate, found from the linearization of Eq. (1), is real, $0.3775$.

### IV. Concluding remarks

We have introduced the 1D setting with the singular modulation of the SF nonlinearity, possibly embedded into the SDF background and combined with the singular linear repulsive potential. This setting, which may be realized in optics and BEC, extends the analysis of solitons in the general class of physical models based on NLSEs. Properties of fundamental solitons, which are fully stable, were investigated analytically and verified numerically. The fundamental soliton modes found here may also emulate solitons of the cubic NLSE in the space of fractional dimension, $0<D<1$. Dipoles (odd solitons) are stable in the semi-infinite domain, where they replace the fundamental solitons.

A natural issue is extension of the present setting into the 2D space. While this is impossible with the cubic nonlinearity, which will inevitably lead to the collapse in 2D, the extension is feasible for the quadratic (second-harmonic-generating, alias $\chi^{(2)}$) nonlinearity. Although detailed analysis of such a model should be a subject of a separate work, it is relevant to write here the underlying equations for amplitudes of the fundamental-frequency and second-harmonic waves, $u(x,y,z)$ and $v(x,y,z)$, where $z$ is the propagation distance, and $(x,y)$ are the transverse coordinates [cf. Eq. (1)]:

$$\begin{cases} iu_z + \dfrac{1}{2}\nabla^2 u + r^{-\alpha} u^* v = 0, \\ 2iv_z + \dfrac{1}{2}\nabla^2 v - Qv + \dfrac{1}{2} r^{-\alpha} u^2 = 0, \end{cases} \quad (11)$$

where $r = \sqrt{x^2 + y^2}$, and $Q$ is a real mismatch parameter [27]. In fact, Eq. (11) is relevant too as a 1D quadratic model, with $r^{-\alpha}$ replaced by $|x|^{-\alpha}$.

Solutions to Eqs. (11) for fundamental solitons with propagation constant $b$ are looked for as $\{u(x,y,z), v(x,y,z)\} = \{e^{ibz}\phi(x,y), e^{2ibz}\psi(x,y)\}$, with real functions $\phi(x,y)$ and $\psi(x,y)$ obeying the stationary equations:

$$\begin{cases} b\phi = \dfrac{1}{2}\nabla^2 \phi + r^{-\alpha}\phi\psi, \\ 4b\psi = \dfrac{1}{2}\nabla^2 \psi - Q\psi + \dfrac{1}{2} r^{-\alpha}\phi^2, \end{cases} \quad (12)$$

cf. Eq. (2). However, the scaling argument, which led to Eq. (3) in the model with the cubic nonlinearity, applies to Eq. (12) solely in the case of zero mismatch, $Q=0$. In that case, the scaling laws for the 2D and 1D versions of system (11) take the following form:

$$\begin{cases} U_{2D}(b) \equiv \iint [\phi^2(x,y) + 4\psi^2(x,y)] dx dy = b^{1-\alpha} U_{2D}(b=1), \\ U_{1D}(b) \equiv \int_{-\infty}^{+\infty} [\phi^2(x) + 4\psi^2(x)] dx = b^{3/2-\alpha} U_{1D}(b=1). \end{cases}$$

The comparison of these relations with Eq. (3) suggests that 2D and 1D solitons, supported by the singularly modulated $\chi^{(2)}$ nonlinearity with the zero mismatch, may exist and be stable for the singularity powers $\alpha_{2D} < 1$ and $\alpha_{1D} < 3/2$, respectively.

On the other hand, nonzero mismatch breaks the scaling invariance and affects the existence and stability areas for solitons. In particular, in the limit of large positive $Q$, the usual cascading-limit approximation may be used in the second equation of system (11), $v \approx (1/2) r^{-\alpha} u^2$ [27]. Upon the substitution into the first equation, this approximation yields the equation with the effective cubic nonlinearity, which is relevant only in the 1D case: $iu_z + (1/2) u_{xx} + (1/2) |x|^{-2\alpha} |u|^2 u = 0$. According to the above results, the latter equation gives rise to stable fundamental solitons at $\alpha < 1/2$. Thus, the transition from $Q=0$ to $Q \to +\infty$ leads to the reduction of the stability limit for the stable fundamental solitons in the 1D $\chi^{(2)}$ system from $\alpha<3/2$ to $\alpha<1/2$. Further analysis of the $\chi^{(2)}$ system will be reported elsewhere.

*Acknowledgments*. We thank Y. V. Kartashov, V. V. Konotop, L. Torner, and V. A. Vysloukh for valuable discussions. The work of O.V.B. was supported by the Ministry of Science and Innovation of Spain, grant FIS2009-09928. B.A.M. appreciates hospitality of ICFO (Institut de Ciencies Fotoniques) at Castelldefels (Barcelona).